\documentclass[11pt,twosided]{article}
\date{}
\usepackage{extsizes}
\usepackage{amsmath}
\usepackage{amsfonts}
\usepackage{amssymb}
\usepackage{graphicx}
\date{\today}
\begin{document}

\title{{\bf{Photon velocity, power spectrum in Unruh effect with modified dispersion relation }}}

\author{
{\bf {\normalsize Arnab Mukherjee}$^{a}
$\thanks{a.m.official.1123@gmail.com}}, 
{\bf {\normalsize Sunandan Gangopadhyay}
$^{b}$\thanks{ sunandan.gangopadhyay@bose.res.in, sunandan.gangopadhyay@gmail.com}},
{\bf {\normalsize Manjari Dutta}
$^{c}$\thanks{manjaridutta@boson.bose.res.in}}\\
$^{a}$ {\normalsize Department of Physics, Jadavpur University, Kolkata 700032, India}\\
$^{b,c}$ {\normalsize Department of Theoretical Sciences},\\
{\normalsize S.N. Bose National Centre for Basic Sciences},\\
{\normalsize JD Block, Sector III, Salt Lake, Kolkata 700106, India}\\
%$^{c}${\normalsize National Institute for Theoretical Physics (NITheP), University of Stellenbosch, Stellenbosch 7600, South Africa}\\
%$^{d}${\normalsize Institute of Theoretical Physics, University of Stellenbosch, Stellenbosch 7600, South Africa}
}
\date{}
\maketitle
\begin{abstract}
\noindent In this paper we propose a new form of generalized uncertainty principle which involves both a linear as well as a quadratic term in the momentum. From this we have obtained the corresponding modified dispersion relation which is compared with the corresponding relation in rainbow gravity. The new form of the generalized uncertainty principle reduces to the known forms in appropriate limits. We then calculate the modified velocity of photons and we find that it is energy dependent, allowing therefore for a superluminal propagation. We then derive the $1+1$-dimensional Klein-Gordon equation taking into account the effects of the modified dispersion relation. The positive frequency mode solution of this equation is then used to calculate the power spectrum arising due to the Unruh effect. The result shows that the power spectrum depends on the energy of the particle owing its origin to the presence of the generalized uncertainty principle. Our results capture the effects of both the simplest form as well as the linear form of the generalized uncertainty principle and also points out an error in the result of the power spectrum up to first order in the generalized uncertainty principle parameter existing in the literature.          

\end{abstract}
\vskip 1cm
%%%%%%%%%%%%%%%%%%%%%%%%%%%%%%%%%%%%%%%%%%%%%%%%%%%%%%%%%%%%%%%%%%

%%%%%%%%%%%%%%%%%%%%%%%%%\section{Introduction  } %%%%%%%%%%%%%%%%%%%

\section{Introduction}

\noindent In 1927 Heisenberg proposed the \textit{Uncertainty Principle} \cite{hberg} from his famous thought experiment. At that time physicists could not believe that the measurement of one physical observable could hamper the measurement of another observable simultaneously. In this principle there is a fundamental limit for the measurement accuracy with which certain pairs of physical observables such as position and momentum or energy and time can be measured simultaneously. Around mid sixties, the concept of fundamental minimum measurable length was introduced \cite{mead}. The existence of this minimum measurable length would mean that Heisenberg uncertainty principle has to be modified. It is for this reason that generalized uncertainty principle (GUP) was introduced in the literature \cite{ama}-\cite{kempf1}. It is observed that the GUP not only has a minimum measurable length but also a maximum measurable momentum \cite{ali}. There has been a lot of investigations thereafter incorporating the effects of the GUP in black hole thermodynamics \cite{adler}-\cite{sg1}, quantum systems such as
particle in a box, Landau levels, simple harmonic oscillator \cite{das}-\cite{das2}.

\noindent In this paper we consider a new form of the GUP which is a combination of both the linear and the simplest forms of the GUP exisiting in the literature. Taking this new form of GUP, we derive the corresponding dispersion relation. From this we obtain the velocity of photons upto second order in the linear GUP parameter and first order in the quadratic GUP parameter. We then derive the Klein-Gordon equation in $(1+1)$-dimensions taking into account the effects of the modified dispersion relation. This equation is then solved in an iterative approach and we work with the positive frequency mode solution only. The solution is then written down for a uniformly accelerating observer using the Rindler coordinate transformations. Using this solution for the uniformly accelerating observer, we derive the emission spectrum arising from the Unruh effect.
Our result capture the effects of both the linear GUP as well as the quadratic GUP parameters in the power spectrum. It also differs from the result in \cite{majhi} upto first order in the linear GUP parameter as the result there is dimensionally incorrect. 

\noindent The paper is organized as follows. In section 2, we consider the new form of the GUP and using this we derive the modified dispersion relation and obtain the photon
velocity. In section 3, we derive the Klein-Gordon equation incorporating the modified dispersion relation and solve this equation. In section 4, we obtain the power spectrum of the Unruh effect using the solution in section 3. We conclude in section 5.

%%%%%%%%%%%%%%%%%%%%%%%%%%%%%%%%%%%%%%%%%%%%%%%%%%%%%%%%%%%%%%%%%%%%%%%%%%%%%%%%

%%%%%%%%%%%%%%%%%%%%%%%%%%%%%%%%%%%%%%%%%%%%%%%%%%%%%%%%%%%%%%%%%%%%%%%%%%%%%%%%

\section{Modified dispersion relation and photon velocity}
In this section our goal is to derive the GUP corrected dispersion relation. Let $p^{A}$ be the modified four momentum and $k^{A}$ be the usual four momentum. Now we introduce a general form of the GUP given by the expression
\begin{equation}
[x_i, p_j]=i\hbar\left[\delta_{ij}-\alpha\left(\delta_{ij}p+\dfrac{p_i p_j}{p}\right)
+\beta\left(\delta_{ij}p^2+2p_i
p_j\right)-\alpha^2(\delta_{ij}p^2+p_i p_j)
\right]\label{5}
\end{equation}
where ${{p}}^2\equiv{\vert\vec{{p}}\vert}^2= \eta_{ij}p^{i}p^{j}$; $i,j=1,2,3$. 
The above commutator ensures by the Jacobi identity that
\begin{equation}
[x_i, x_j]=0~,~[p_i, p_j]=0.\label{Jac}
\end{equation}
The commutation relation between the position operator and the usual three momentum operator is the standard Heisenberg algebra given by
\begin{equation}
[x_i, k_j]=i\hbar\delta_{ij}.\label{5H}
\end{equation}
The relations between the modified and the usual momenta which gives the commutation relation (\ref{5}) read 
\begin{align}
\label{1}
p^{0}&=k^{0}\\
p^{i}&=k^{i}(1-\alpha{k}+\beta{{k}}^2)\label{2}
\end{align}
where $\alpha=\alpha_{0}/(M_{Pl}c)$ and $\beta=\beta_{0}/(M_{Pl}c)^2$ are small parameters, $M_{Pl}$ is the Planck mass, ${{k}}^2\equiv{\vert\vec{{k}}\vert}^2= \eta_{ij}k^{i}k^{j}$.

\noindent It is reassuring to note that in the limit $\alpha\rightarrow 0$, 
eq.(\ref{5}) reduces to the simplest form of the GUP proposed in the literature \cite{kempf}
\begin{equation}
[x_i, p_j]=i\hbar\left[\delta_{ij}+\beta\left(\delta_{ij}p^2+2p_i p_j\right)\right]\,.\label{2a}
\end{equation}
Further, setting $\beta=2\alpha^2$ in eq.(\ref{5}) 
yields the expression \cite{das}
\begin{equation}
[x_i, p_j]=i\hbar\left[\delta_{ij}-\alpha\left(\delta_{ij}p+\dfrac{p_i p_j}{p}\right)+\alpha^2\left(\delta_{ij}p^2+3p_i
p_j\right)\right]\,.\label{2b}
\end{equation}
Setting $i=j$, eq.(\ref{5}) leads to the following modified uncertainty relation 
\begin{equation}
\Delta x_i \Delta p_i\geq \frac{\hbar}{2}\left(1-\alpha \left\langle p+\frac{p_i p_i}{p}\right\rangle-(\alpha^2-\beta)\left((\Delta p)^2+{\langle p\rangle}^2\right)-(\alpha^2-2\beta)\left(({\Delta p_i})^2+{\langle p_i\rangle}^2\right)\right).\label{2c}
\end{equation}
We now consider the background spacetime metric $\eta_{AB},\,\, (A,B=0,1,2,3)$ in (3+1)-dimensions to be the Minkowski spacetime with signature $(-,+,+,+)$, that is
\begin{equation}
ds^2=\eta_{AB}dx^{A}dx^{B}= -c^2dt^2+\eta_{ij}dx^{i}dx^{j}
\end{equation}
with $\eta_{00}=-1$, $\eta_{ij}=\delta_{ij}$.
Hence the square of the four momentum in this background becomes
\begin{align}
p^Ap_{A}&=\eta_{AB}p^Ap^B\nonumber\\&=-(p^0)^2+\eta_{ij}p^{i}p^{j}\nonumber\\
&=-(k^0)^2+\eta_{ij}k^{i}k^{j}(1-\alpha{k}+\beta{{k}}^2)^2\nonumber\\
&=-(k^0)^2+{{k}}^2(1-\alpha{k}+\beta{{k}}^2)^2
\end{align}
where eq.(s)(\ref{1}, \ref{2}) have been used in the third line of the equality.

\noindent Keeping terms up to $\mathcal{O}({\alpha}^2,\,\beta)$ in the above expression, we obtain
\begin{align}
p^Ap_{A}&=-(k^0)^2+{{k}}^2[1-2\alpha{k}+{\alpha}^2{{k}}^2+2{\beta}{{k}}^2]\nonumber\\
&=k^Ak_A+{{k}}^2[-2\alpha{k}+{\alpha}^2{{k}}^2+2{\beta}{{k}}^2].\label{6}
\end{align}
Using the usual dispersion relation
\begin{equation}
k^Ak_A=-m^2c^2
\end{equation}
eq.(\ref{6}) takes the form
\begin{equation}
p^Ap_{A}=-m^2c^2+{{k}}^2[-2\alpha{k}+({\alpha}^2+2\beta){{k}}^2].\label{16}
\end{equation}
Setting $\beta=2\alpha^2$, the above relation reduces to \cite{majhi}
\begin{equation}
p^Ap_{A}=-m^2c^2+{{k}}^2[-2\alpha{k}+5{\alpha}^2{{k}}^2].\label{16a}
\end{equation}
To express the usual momentum $k$ in terms of the modified momentum $p$ up to $\mathcal{O}({\alpha}^2,\,\beta)$, we choose an ansatz of the form
\begin{equation}
{k}=a_{0}+\alpha a_1 +\beta a_{2}+{\alpha}^2a_{3}.\label{seven}
\end{equation}
Now taking the magnitude of both sides of eq.(\ref{2}), we get
\begin{equation}
p=k(1-\alpha{k}+\beta{{k}}^2),\,p=\vert \vec{p}\vert.\label{mag}
\end{equation}
Substituting eq.(\ref{seven}) in eq.(\ref{mag}), we get
\begin{equation}
\text{p}= a_{0}+(a_{1}-a_{0}^2)\alpha+(a_{3}-2a_{0}a_{1}){\alpha}^2+(a_{0}^3+a_2)\beta.
\end{equation}
Comparing coefficients of ${\alpha}^0,\,\alpha,\,{\alpha}^2,\,\beta$ 	
on both sides of the above equation, we have
\begin{eqnarray}
a_0=p\\
a_1=a_{0}^2=p^2\\
a_3=2a_0a_1=2p^3\\
a_2=-a_{0}^3=-p^3.
\end{eqnarray}
This gives the following relation between the usual three momentum $k^i$ in terms of the modified three momentum $p^i$
\begin{equation}
k^i=p^i[1+\alpha{{p}}+(2{\alpha}^2-\beta){{p}}^2].
\end{equation}
Substituting the above expression for the usual three momentum $k^i$ in eq.(\ref{16}) and retaining terms up to $\mathcal{O}({\alpha}^2,\,\beta)$, the modified dispersion relation corresponding to the form of the GUP considered here takes the form 
\begin{eqnarray}
p^Ap_{A}=-m^2c^2+{{p}}^2[-\:2\alpha{p}-(5{\alpha}^2-2\beta){{p}}^2].\label{20}
\end{eqnarray}
Writing the left hand side of the above equation as
\begin{equation}
p^Ap_{A}=-\mathcal{M}^2c^2 
\end{equation}
yields
\begin{equation}
\mathcal{M}=\sqrt{m^2+\frac{{{p}}^2[\:2\alpha{p}+(5{\alpha}^2-2\beta){{p}}^2]}{c^2}}\,.
\end{equation} 
The above result indicates that $\mathcal{M}$ is an effective mass of the particle generating solely due to the GUP.\\ 
From eq.(\ref{20}), the time component of the four momentum squared, upto $\mathcal{O}({\alpha}^2,\,\beta)$ can be written as
\begin{eqnarray}
(p^0)^2=m^2c^2+
{{p}}^2[1+2\alpha{p}+(5{\alpha}^2-2\beta){{p}}^2].
\end{eqnarray}
Hence the energy of the particle upto $\mathcal{O}({\alpha}^2,\,\beta)$ is given by
\begin{eqnarray}
E^2=m^2c^4+{{p}}^2c^2[1+2\alpha{p}+(5{\alpha}^2-2\beta){{p}}^2].\label{d1}
\end{eqnarray}
The above relation is the most general modified dispersion relation. It is interesting to compare this result with the rainbow gravity generalization of
the modified dispersion relations in doubly special relativity to curved
spacetime. These modified dispersion relations are given
by \cite{smolin1, smolin2}
\begin{eqnarray}
E^2 f^{2}(E/E_p)-p^2 c^2 g^{2}(E/E_p)=m^2 c^4\label{smolin1}
\end{eqnarray}
where $E_p$ is the Planck energy and the functions $f(E/E_p)$ and $g(E/E_p)$ are called rainbow functions. Specific forms of the rainbow functions read \cite{cam1}
\begin{eqnarray}
f(E/E_p)=1~,~g(E/E_p)=\sqrt{1-\eta\left(\frac{E}{E_p}\right)^n}\label{rainbow}
\end{eqnarray}
where $\eta$ is the rainbow parameter. In \cite{sgepl}, it was argued from the universality of logarithmic corrections to black hole entropy that $n$ gets restricted to $n=1,2$. Setting $n=2$, eq.(\ref{smolin1}) gives
\begin{eqnarray}
E^2=\frac{m^2 c^4+p^2 c^2}{1+\eta\frac{p^2 c^2}{E_{p}^2}}~.\label{rbow1}
\end{eqnarray}
Keeping terms upto linear order in $\eta p^2 c^2/E_{p}^2$ in the above relation yields
\begin{eqnarray}
E^2=m^2 c^4+p^2 c^2 \left[1-\eta\frac{m^2 c^4}{E_{p}^2}-\eta\frac{p^2 c^2}{E_{p}^2}\right]~.\label{rbow2}
\end{eqnarray}
The above relation has a very similar structure with the one derived in eq.(\ref{d1}).

\noindent Setting $\beta=2\alpha^2$ in eq.(\ref{d1}), we get
\begin{equation}
E^2=m^2c^4+{{p}}^2c^2[1+2\alpha{p}+{\alpha}^2{{p}}^2].\label{d2}
\end{equation} 
It is reassuring to note that in the absence of quantum gravity corrections, that is, $\alpha=0$ and $\beta=0$, we have $p^i=k^i$, and hence one gets back the standard dispersion
\begin{equation}
 E^2=m^2c^4+{{k}}^2 c^2. \label{e1}
\end{equation}
We shall now proceed to investigate how the velocity of photon gets affected by the modified dispersion relation (eq.(\ref{d1})). In the usual case, the velocity of photon $c=E/\text{k}$. However, since the momentum is modified here due to the GUP, hence it is expected that the velocity of photon will also get modified. In Minkowskian background, the velocity of photon can be calculated as 
\begin{equation}
u=\frac{{\partial}E}{{\partial}{p}}\,.\label{onefour}
\end{equation} 
Setting $m=0$ in eq.(\ref{d1}), we get
\begin{equation}
E^2={{p}}^2c^2[1+2\alpha{p}+(5{\alpha}^2-2\beta){{p}}^2].\label{onefive}
\end{equation}
Substituting eq.(\ref{onefive}) in eq.(\ref{onefour}) and keeping terms up to $\mathcal{O}(\alpha^2,\,\beta)$, we get
\begin{equation}
u=\frac{{\partial}E}{{\partial}{p}}=c[1+2\alpha{p}+3(2\alpha^2-\beta){p}^2].\label{onesix}
\end{equation}
Now we proceed to invert eq.(\ref{onefive}) to express $p$ in terms of energy $E$. This is required to rewrite eq.(\ref{onesix}) in terms of energy $E$. To do this we take $p$ to be of the form
\begin{equation}
{p}=a+\alpha{b}+\beta{e}+\alpha^2{d}+\mathcal{O}(\alpha^3,\beta^2,\alpha^2\beta).          \label{oneseven}
\end{equation}
Substituting eq.(\ref{oneseven}) in eq.(\ref{onefive}) and comparing the coefficients of ${\alpha}^0,\,\alpha,\,\beta,\,{\alpha}^2$ on both sides of the equation, we get
\begin{align}
a^2c^2&=E^2\\
abc^2+c^2a^3&=0\\
ac^2e-a^4c^2&=0\\
b^2c^2+2adc^2+6c^2a^2b+5c^2a^4&=0\,.
%a^2c^{'2}=0\\
%a^3c^2+abc^2=0\\
%abc^2-a^4c^2=0\\
%6a^2bc^2+5a^4c^2+2adc^2+b^2c^2=0
\end{align}
Solving the above equations, we obtain
\begin{align}
a&=\dfrac{E}{c}\\
b&=-a^2=-\dfrac{E^2}{c^2}\\
d&=0\\
e&=a^3=\dfrac{E^3}{c^3}\,.
\end{align}
Substituing the above values in eq.(\ref{oneseven}), we have
\begin{equation}
p=\frac{E}{c}\left[1-\alpha\frac{E}{c}+\beta{\frac{E^2}{c^2}}\right]+\mathcal{O}(\alpha^3,\beta^2,\alpha^2\beta).\label{twotwo}
\end{equation}
Substituting eq.(\ref{twotwo}) into eq.(\ref{onesix}) and keeping terms up to $\mathcal{O}(\alpha^2,\,\beta)$, the modified photon velocity is obtained to be
\begin{equation}
u=c\left[1+\frac{2{\alpha}E}{c}+\frac{(4\alpha^2-3\beta)E^2}{c^2}\right].\label{e2}
\end{equation}
The above result for the velocity of photon captures the effect of the GUP for both the linear as well as the quadratic terms in momentum in the GUP (\ref{2c}).

\noindent Setting $\beta=2\alpha^2$ in the above relation, we get
\begin{equation}
u=c\left[1+\frac{2{\alpha}E}{c}-\frac{2\alpha^2{E^2}}{c^2}\right].\label{e3}
\end{equation}
From the above relation, we observe that the velocity of photon is energy dependent. This energy dependence is due to the modified dispersion relation arising from the GUP. The other important point to note is that the photon velocity is larger than the speed of light $c$ which indicates that quantum gravity effects allow a superluminal photon propagation.

%%%%%%%%%%%%%%%%%%%%%%%%%%%%%%%%%%%%%%%%%%%%%%%%%%%%%%%%%%
%%%%%%%%%%%%%%%%%%%%%%%%%%%%%%%%%%%%%%%%%%%%%%%%%%%%%%%%%%

\section{Klein-Gordon equation and modified dispersion relation}
In this section we are going to write down the modified Klein-Gordon equation in $(1+1)$-dimensional Minkowski spacetime
\begin{eqnarray}
ds^2=-c^2dT^2+dX^2\quad.
\end{eqnarray}
To obtain the modified Klein-Gordon equation, we first recast eq.(\ref{6}) in the form
\begin{equation}
{p_A{p^A}}=-{(k^0)}^2+k^2+k^2[-2{\alpha}k+\alpha^2{k^2}+2{\beta}k^2].
\end{equation}
Elevating $k^0$ and $k$ to operators and using their standard representations
\begin{eqnarray}
k^0=\dfrac{i\hbar}{c}\partial_{T}\,\,\,\,,\,\,\,\,k^1=-i\hbar\partial_X
\end{eqnarray}
and keeping terms up to $\mathcal{O}({\alpha}^2,\,\beta)$, we get the modified Klein-Gordon equation in $(1+1)$-dimensions for massless particles to be
\begin{align}
&{p_A{p^A}}\,\,\Phi(T,X)=\hbar^2\left[\dfrac{1}{c^2}\partial_T^2-\partial_X^2-2i\alpha\hbar\partial_X^3+(\alpha^2+2\beta)\hbar^2\partial_X^4\right]\Phi(T,X)=0.
\end{align}
The third and fourth terms in the right hand side of the above equation are the ones that have emerged due to the GUP. We now take a solution of the above equation in the form
\begin{equation}
{\Phi(T,X)}=\exp(-i{\omega}T)\Psi(X)\,\label{kg1}
\end{equation} 
where $\hbar\omega$ is the energy of the particle.

\noindent Substituting this in the above equation, we get
\begin{align}
\left[\partial_X^2+2i\alpha\hbar\partial_X^3-(\alpha^2+2\beta)\hbar^2\partial_X^4
+\dfrac{\omega^2}{c^2}\right]\Psi(X)=0\,.\label{kg2}
\end{align}
The above equation reduces to that derived in \cite{majhi} upto $\mathcal{O}(\alpha, \beta^{0})$.

\noindent Setting
\begin{equation}
\Psi(X)=\exp(nX)\label{kg3}
\end{equation}
 we obtain
\begin{align*}
n^2+2i\hbar\alpha{n^3}-(\alpha^2+2\beta)\hbar^2{n^4}+\dfrac{\omega^2}{c^2}=0\,.\label{kg4}
\end{align*}
To solve this equation, we choose an ansatz 
\begin{equation}
n=n_0+\alpha{n_1}+\beta{n_2}+{\alpha^2}n_3\label{kg5}
\end{equation}
where $n_0,n_1,n_2,n_3$ are to be determined. Substituting this ansatz in the above equation and comparing the coeeficients of $\alpha^0,\alpha,\beta$ and $\alpha^2$ on both sides of the equation, we get
\begin{align*}
\dfrac{\omega^2}{c^2}+n_0^2=0&\\
2n_0{n_1}+2i\hbar{n_0^3}=0&\\
{n_1^2+2n_0{n_3}+6i\hbar{n_0^2}{n_1}}-\hbar^2 n_{0}^4=0&\\
-2\hbar^2{n_0^4}+2n_0{n_2}=0&. \label{kg6}
\end{align*}
Solving these equations, we get 
\begin{align}
n_0&=\dfrac{i\omega}{c}\\
n_1&=i\hbar\dfrac{\omega^2}{c^2}\\
n_2&=-i\hbar^2\dfrac{\omega^3}{c^3}\\
n_3&=2i\hbar^2\dfrac{\omega^3}{c^3}\quad. \label{kg7}
\end{align}
Note that we have considered only the outgoing mode solutions since they give the radiation spectrum. The positive frequency outgoing solution of the modified Klein-Gordon equation therefore reads
%we get \,\,\,\,\,\,\,\,\,\,\,\,\,\,\,\,$n_2$=$-i\hbar^2\dfrac{\omega^3}{c^3}$\\
%Now as the Rindler horizon trap the ingoing mode, therefore ingoing solutions will not be considered. Outgoing modes are escaped from the horizon and acts as a source of radiation comes from the horizon. As here we are discussing about the radiation spectrum, measured by the uniformly accelerated observer, therefore we will consider only outgoing mode solutions, though n has four solutions mathematically. We need only one solution arised from $n_0=\dfrac{i\omega}{c}$. So,the root is
%\begin{equation}
%n^{(1)}=\dfrac{i\omega}{c}+i\alpha\hbar\dfrac{\omega^2}{c^2}+i(2\alpha^2-\beta)\hbar^2\dfrac{\omega^3}{c^3}
%\end{equation}
 %The first solution $n^{(1)}$ is the only solution that leads to positive spatial momentum, and thus it is only that will be considered in the following analysis. So putting the value of n, the positive frequency outgoing solution of the modified Klein-Gordon equation becomes,
\begin{align}
\Phi(T,X)=exp\left[-i\omega\left(T-\dfrac{X}{c}\right)+i\alpha\dfrac{\hbar\omega^2}{c^2}X+i(2\alpha^2-\beta)\dfrac{\hbar^2\omega^3}{c^3}X\right].\label{threefive}
\end{align}
The above solution reduces to that derived in \cite{majhi} upto $\mathcal{O}(\alpha, \beta^{0})$. Interestingly, the above solution also reduces to that derived in \cite{majhi} for $\beta=2\alpha^2$. In the next section, we shall use this solution to investigate the power spectrum of a uniformly accelerated observer.

%%%%%%%%%%%%%%%%%%%%%%%%%%%%%%%%%%%%%%%%%%%%%%%%%%%%%%%%%%%%%%%
%%%%%%%%%%%%%%%%%%%%%%%%%%%%%%%%%%%%%%%%%%%%%%%%%%%%%%%%%%%%%%%

\section{GUP corrected power spectrum in Unruh effect}
In this section we essentially follow the analysis in \cite{paddy} to compute the power spectrum of a uniformly accelerating observer.
We first consider a uniformly accelerated frame known as Rindler frame. The coordinate transformation equations connecting Minkowski and Rindler frames with respect to an observer in Rindler frame moving along the $x$-axis read
\begin{eqnarray}
X(\tau)=\frac{c}{\kappa}\cosh(\kappa\tau)\,\,\,\,\,\,,\,\,\,\,\,T(\tau)=\frac{1}{\kappa}\sinh(\kappa\tau)\label{u1}
\end{eqnarray} where $\tau$ is the proper time of the uniformly accelerating observer. Hence the wave function (eq.(\ref{threefive})) as seen by the Rindler observer will have the form
%\begin{equation*}
%T({\tau})-\dfrac{X(\tau)}{c}=\dfrac{1}{\kappa}\left[sinh(\kappa\tau)-cosh(\kappa\tau) \right]
%\end{equation*}
%Now
\begin{align}
\phi{[T({\tau}),X({\tau})]}=exp\left[\dfrac{i\omega}{\kappa}e^{-\kappa\tau}\left(1+\dfrac{\alpha\hbar\omega}{2c}+\dfrac{(2\alpha^2-\beta)\hbar^2\omega^2}{2c^2}\right)\right]exp\left[\dfrac{i\hbar\omega^2}{2c\kappa}e^{\kappa\tau}\left(\alpha+(2\alpha^2-\beta)\dfrac{\hbar\omega}{c} \right) \right]\,.\label{43}
\end{align}
The above relation can be obtained by substituing eq.(\ref{u1}) in eq.(\ref{threefive}).

\noindent The power spectrum is now given by \cite{paddy}
\begin{equation}
P(\nu)={\vert f(\nu)\vert}^2\label{u1a}
\end{equation} 
where $f(\nu)$ is the Fourier transform of $\phi(\tau)$
\begin{equation}
f(\nu)=\int_{-\infty}^{+\infty}{d{\tau}}\,{\Phi(\tau)}e^{i{\nu}\tau}\quad.\label{u2}
\end{equation}
Substituting eq.(\ref{43}) in the above relation yields
\begin{align}
f(\nu)=\int_{-\infty}^{+\infty}{d{\tau}}\,&exp\left[\dfrac{i\omega}{\kappa}e^{-\kappa\tau}\left(1+\dfrac{\alpha\hbar\omega}{2c}+\dfrac{(2\alpha^2-\beta)\hbar^2\omega^2}{2c^2}\right)\right]\nonumber\\
&\times \,\,exp\left[\dfrac{i\hbar\omega^2}{2c\kappa}e^{\kappa\tau}\left(\alpha+(2\alpha^2-\beta)\dfrac{\hbar\omega}{c} \right) \right]
 e^{i{\nu}\tau}\,.\label{u3}
\end{align}
Now expanding the second term of the above equation and keeping terms upto $\mathcal{O}({\alpha}^2,\beta)$, we obtain
%Let us denote the each terms of the integral as

\begin{align}
f(\nu)&=\int^{\infty}_{-\infty}\left[ exp\left[\dfrac{i\omega}{\kappa}e^{-\kappa\tau}\left(1+\dfrac{\alpha\hbar\omega}{2c}+\dfrac{(2\alpha^2-\beta)\hbar^2\omega^2}{2c^2}\right)\right]    +i\nu\tau\right]d\tau \nonumber\\
&+\int^{\infty}_{-\infty}\left[ \dfrac{i\hbar\omega^2}{2c\kappa}\left(\alpha+(2\alpha^2-\beta)\dfrac{\hbar\omega}{c} \right)exp\left[i\nu\tau+\kappa\tau+\dfrac{i\omega}{\kappa}e^{-\kappa\tau}\left(1+\dfrac{\alpha\hbar\omega}{2c}+\dfrac{(2\alpha^2-\beta)\hbar^2\omega^2}{2c^2}\right) \right]   \right]d\tau \nonumber\\
&-\int^{\infty}_{-\infty}\dfrac{\alpha^2\omega^4\hbar^2}{8c^2\kappa^2}exp\left[i\nu\tau+2\kappa\tau+\dfrac{i\omega}{\kappa}e^{-\kappa\tau}\left(1+\dfrac{\alpha\hbar\omega}{2c}+\dfrac{(2\alpha^2-\beta)\hbar^2\omega^2}{2c^2}\right) \right]d\tau\,.
\label{fivefive}
\end{align}
To compute the above integrals we need to rewrite the integral in a suitable form. For doing that we introduce a new variable $v=e^{\kappa\tau}$. Using this, 
eq.(\ref{fivefive}) takes the form
\begin{align}
f(\nu)&=\int^{\infty}_{0}\dfrac{1}{\kappa}v^{-\left(1+\dfrac{i\nu}{\kappa}\right)}exp\left[\dfrac{i\omega{v}}{\kappa}\left(1+\dfrac{\alpha\hbar\omega}{2c}+\dfrac{(2\alpha^2-\beta)\hbar^2\omega^2}{2c^2}\right)\right]dv \\ \nonumber
&+\int^{\infty}_{0}\dfrac{i\hbar\omega^2}{2c\kappa^2}\left(\alpha+(2\alpha^2-\beta)\dfrac{\hbar\omega}{c} \right)v^{-\left(1+\dfrac{i\nu}{\kappa}\right)-1}exp\left[\dfrac{i{\omega}Av}{\kappa}\right]dv \\ \nonumber
&-\dfrac{\alpha^2\omega^4\hbar^2}{8c^2\kappa^3}\int^{\infty}_{0}v^{-\left(2+\dfrac{i\nu}{\kappa}\right)-1}exp\left[\dfrac{i{\omega}Av}{\kappa}\right]dv\label{u4}
\end{align}
where
\begin{equation} A=\left(1+\frac{\alpha\hbar\omega}{2c}+\dfrac{(2\alpha^2-\beta)\hbar^2\omega^2}{2c^2}\right)\,.\label{u4a}
\end{equation} 
To perform the above integrals, we use the standard integral
\begin{equation}
\int^{\infty}_{0}x^{s-1}exp(-bx)dx=exp(-s\,lnb)\Gamma(s)\,\,.\label{u5}
\end{equation}
Using this and keeping the terms upto $\mathcal{O}({\alpha}^2,\,\beta)$, we obtain 
\begin{align}
f(\nu)=\dfrac{1}{\kappa}\left[\dfrac{wA}{\kappa}\right]^{\dfrac{i\nu}{\kappa}}exp\left(\dfrac{\pi\nu}{2\kappa}\right)\Gamma\left(-\dfrac{i\nu}{\kappa}\right)&\left[1-\dfrac{\alpha\hbar{\omega^3}}{2c{\kappa^2}\left(1+\dfrac{i\nu}{\kappa}\right) }+\frac{\beta\hbar^2\omega^4}{2c^2{\kappa^2}\left(1+\dfrac{i\nu}{\kappa}\right)}-\frac{5\alpha^2\hbar^2\omega^4}{4c^2{\kappa^2}\left(1+\dfrac{i\nu}{\kappa}\right)}\right.\nonumber\\
&+\left.\dfrac{\hbar^2\alpha^2\omega^6}{8c^2\kappa^4\left(1+\dfrac{i\nu}{\kappa}\right)\left(2+\dfrac{i\nu}{\kappa}\right) }
\right].\label{u7}
\end{align}
Hence the power spectrum with a negative frequency is given by
\begin{align}
{\vert{f(-\nu)}\vert}^2=\frac{2\pi}{\nu\kappa}\frac{1}{(e^{\frac{2\pi\nu}{\kappa}}-1)}\left[1-\frac{\alpha\hbar\omega^3}{c\kappa^2(\frac{\nu^2}{\kappa^2}+1)}+\frac{\beta\hbar^2\omega^4}{c^2\kappa^2(\frac{\nu^2}{\kappa^2}+1)}-\frac{5\;\alpha^2\hbar^2\omega^4}{2c^2\kappa^2(\frac{\nu^2}{\kappa^2}+1)}+\frac{3\alpha^2\hbar^2\omega^6}{2c^2{\kappa^4(\frac{\nu^2}{\kappa^2}+1)}{(\frac{\nu^2}{\kappa^2}+4)}}\right]\,.\label{u8}
\end{align}
From this we can obtain the power spectrum per logarithmic band to be
\begin{equation}
\nu{\vert{f(-\nu)}\vert}^2=\frac{2\pi}{\kappa}\frac{1}{(e^{\frac{2\pi\nu}{\kappa}}-1)}\left[1-\frac{\alpha\hbar\omega^3}{c\kappa^2(\frac{\nu^2}{\kappa^2}+1)}+\frac{\beta\hbar^2\omega^4}{c^2\kappa^2(\frac{\nu^2}{\kappa^2}+1)}-\frac{5\;\alpha^2\hbar^2\omega^4}{2c^2\kappa^2(\frac{\nu^2}{\kappa^2}+1)}+\frac{3\alpha^2\hbar^2\omega^6}{2c^2{\kappa^4(\frac{\nu^2}{\kappa^2}+1)}{(\frac{\nu^2}{\kappa^2}+4)}}\right]\,.\label{fourfive1}
\end{equation}
A few observations are in place now. 
We would first like to point out that the power spectrum becomes dependent on the frequency $\omega$ and hence the energy of the particle due to the presence of the GUP. This result is in contrast to the result in the absence of the GUP where the power spectrum is independent of the energy of the particle. 
We would then like to point out that upto $\mathcal{O}(\alpha)$ our result does not agree with that obtained in \cite{majhi}, and that there is an error in the result in \cite{majhi} in the power of $\kappa$. 

\noindent Further, setting $\beta=2\alpha^2$ in the above relation gives
\begin{equation}
\nu{\vert{f(-\nu)}\vert}^2=\frac{2\pi}{\kappa}\frac{1}{(e^{\frac{2\pi\nu}{\kappa}}-1)}\left[1-\frac{\alpha\hbar\omega^3}{c\kappa^2(\frac{\nu^2}{\kappa^2}+1)}-\frac{\;\alpha^2\hbar^2\omega^4}{2c^2\kappa^2(\frac{\nu^2}{\kappa^2}+1)}+\frac{3\alpha^2\hbar^2\omega^6}{2c^2{\kappa^4(\frac{\nu^2}{\kappa^2}+1)}{(\frac{\nu^2}{\kappa^2}+4)}}\right]\,.\label{fourfive10}
\end{equation}
Upto $\mathcal{O}(\alpha)$, the above result reduces to
\begin{equation}
\nu{\vert{f(-\nu)}\vert}^2=\frac{2\pi}{\kappa}\frac{1}{(e^{\frac{2\pi\nu}{\kappa}}-1)}\left[1-\frac{\alpha\hbar\omega^3}{c\kappa^2(\frac{\nu^2}{\kappa^2}+1)}\right]\,.\label{fourfive100}
\end{equation}
From the non-negativity of the power spectrum, the above result imposes a constraint on the linear GUP parameter $\alpha$, which reads
\begin{equation}
\frac{\alpha\hbar\omega^3}{c\kappa^2(\frac{\nu^2}{\kappa^2}+1)}<1.\label{constraint}
\end{equation}
Setting $\alpha=0$ in eq.(\ref{fourfive1}), we obtain the power spectrum per logarithmic band for the simplest form (quadratic) of the GUP to be
\begin{equation}
\nu{\vert{f(-\nu)}\vert}^2=\frac{2\pi}{\kappa}\frac{1}{(e^{\frac{2\pi\nu}{\kappa}}-1)}\left[1+\frac{\beta\hbar^2\omega^4}{c^2\kappa^2(\frac{\nu^2}{\kappa^2}+1)}\right]\,.\label{fourfive12}
\end{equation}
Reassuringly we recover the standard result in the limit $\alpha,\,\beta\rightarrow\,0$
\cite{paddy}
\begin{equation}
\nu{\vert{f(-\nu)}\vert}^2=\frac{2\pi}{\kappa}\frac{1}{(e^{\frac{2\pi\nu}{\kappa}}-1)}\,.\label{u9}
\end{equation} 
This indeed shows that the power spectrum is independent of the energy of the particle.

%%%%%%%%%%%%%%%%%%%%%%%%%%%%%%%%%%%%%%%%%%%%%%%%%%%%%%%%%%%%%%%%%%%%%%%%%%%%%%%%%%%%
%%%%%%%%%%%%%%%%%%%%%%%%%%%%%%%%%%%%%%%%%%%%%%%%%%%%%%%%%%%%%%%%%%%%%%%%%%%%%%%%%%%%

\section{Conclusions}

In this paper we have proposed a new form of generalized uncertainty principle which contains both the linear as well as the quadratic terms in the momentum, from which we derive the corresponding modified dispersion relation. We compare this with the dispersion relation in rainbow gravity and observe that both have a very similar structure. The new form of the generalized uncertainty principle reduces to the known forms existing in the literature in appropriate limits. The modified velocity of photons is then obtained from the modified dispersion relation and shows that it is energy dependent, and hence allows for a superluminal propagation. We then derive the $1+1$-dimensional Klein-Gordon equation taking into account the effects of the modified dispersion relation. Solving this equation iteratively, we obtain the power spectrum arising due to the Unruh effect. 
The result shows that the power spectrum depends on the energy of the particle owing its origin to the presence of the generalized uncertainty principle. This implies that the energy of the particle provides a back reaction effect due to the generalized uncertainty principle. This is similar to the back reaction effects observed in rainbow gravity \cite{magueijo}.
Keeping terms upto leading order in the linear generalized uncertainty principle parameter in the power spectrum result, we observe   
that a constraint gets imposed on this parameter. Our results capture the effects of both the simplest form as well as the linear form of the generalized uncertainty principle.

%%%%%%%%%%%%%%%%%%%%%%%%%%%%%%%%%%%%%%%%%%%%%%%%%%%%%%%%%%%%%%%%%%%%%%%%%

\section*{Acknowledgements}SG would like to thank the Inter University Centre for Astronomy and Astrophysics (IUCAA), Pune for the Visiting Associateship.

%%%%%%%%%%%%%%%%%%%%%%%%%%%%%%%%%%%%%%%%%%%%%%%%%%%%%%%%%%%%%%%%%%%%%%%%%


\begin{thebibliography}{99}
\bibitem{hberg} Heisenberg, W, Zeitschrift f\"ur Physik., (1927), \textbf{43} (3-4), 172-198
\bibitem{mead} C.A. Mead, Phys. Rev. 135 (1964) B849.
\bibitem {ama} D. Amati, M. Ciafaloni, G. Veneziano, Phys. Lett. B 216 (1989) 41.
\bibitem{maggi} M. Maggiore, Phys. Lett. B 304 (1993) 65, arXiv:hep-th/9301067.
\bibitem{maggi1} M. Maggiore, Phys. Rev. D 49 (1994) 5182, arXiv:hep-th/9305163.
\bibitem{maggi2} M. Maggiore, Phys. Lett. B 319 (1993) 83, arXiv:hep-th/9309034.
\bibitem{kempf} A. Kempf, G. Mangano, R. B. Mann, Phys. Rev.D 52 (1995) 1108; arXiv:hep-th/9412167.
\bibitem{kempf1} A. Kempf, J. Phys. A 30 (1997) 2093 arXiv:hep-th/9604045.
\bibitem{ali} A.F. Ali, S. Das, E.C. Vagenas, Phys. Lett. B 678 (2009) 497, arXiv:0906.5396[hep-th].


\bibitem{adler}R.J.~Adler, P. Chen, D.I. Santiago, Gen. Rel. Grav. 33 (2001) 2101. 
\bibitem{sg}S.~Gangopadhyay, A.~Dutta, A.~Saha, Gen. Rel. Grav. 46 (2014) 1661.
\bibitem{sg1}S.~Gangopadhyay, A. Dutta, Gen. Rel. Gravit. 46 (2014) 1661.

%\bibitem{brau} F. Brau, J. Phys. A 32 (1999) 7691 arXiv:quant-ph/9905033.
\bibitem{das} S. Das, E.C. Vagenas, Phys. Rev. Lett. 101 (2008) 221301, arXiv:0810.5333 [hep-th];\\
S. Das, E.C. Vagenas, Phys. Rev. Lett. 104 (2010) 119002, arXiv:1003.3208 [hep-th]
\bibitem{das1} S. Das, E.C. Vagenas, A.F. Ali Phys. Lett. B 690 (2010) 407, arXiv:1005.3368 [hep-th].
\bibitem{das2} S. Das, E.C. Vagenas, Can. J. Phys. 87 (2009) 233, arXiv:0901.1768 [hep-th];\\
A.F. Ali, S. Das, E.C. Vagenas, arXiv:1001.2642 [hep-th];\\
S. Basilakos, S. Das, E.C. Vagenas, JCAP 1009 (2010) 027, arXiv:1009.0365 [hep-th];\\
A.F. Ali, S. Das, E.C. Vagenas, Phys. Rev. D 84 (2011) 044013, arXiv:1107.3164 [hep-th].

\bibitem{majhi} B.R. Majhi, E.C. Vagenas, Phys. Lett. B 725 (2013) 477, arXiv:1307.4195v2[gr-qc];

%\bibitem{camelia} G. Amelino-Camelia, Living Rev. Rel. 16 (2013) 5, arXiv:0806.0339 [gr-qc].
%\bibitem{cam1} G. Amelino-Camelia, J.R. Ellis, N.E. Mavromatos, D.V. Nanopoulos, S. Sarkar, Nature 393 (1998) 763 arXiv:astro-ph/9712103.
%\bibitem{rinaldi} M. Rinaldi, Phys. Rev. D 77 (2008) 124029, arXiv:0802.0618 [gr-qc].
%\bibitem{gutti} S. Gutti, S. Kulkarni, L. Sriramkumar, Phys. Rev. D 83 (2011) 064011, arXiv:1005.1807 [gr-qc].
%\bibitem{hari} E. Harikumar, A.K. Kapoor, R. Verma, Phys. Rev. D 86 (2012) 045022, arXiv:1206.6179 [hep-th].
%\bibitem{hari1} E. Harikumar, R. Verma, Mod. Phys. Lett. A 28 (2013) 1350063, arXiv:1211.4304 [hep-th].

\bibitem{smolin1}J.~Magueijo J, L.~Smolin, Phys. Rev. Lett. 88 (2002) 190403; Phys. Rev. D 67 (2003) 044017.
\bibitem{smolin2}J.~Magueijo, L.~Smolin, Phys. Rev. Lett. 88 (2002) 190403.


\bibitem{cam1} G. Amelino-Camelia, J.R. Ellis, N.E. Mavromatos, D.V. Nanopoulos, S. Sarkar, Nature 393 (1998) 763.
\bibitem{sgepl}S.~Gangopadhyay, A.~Dutta, Euro. Phys. Lett. 115 (2016) 50005.

\bibitem{paddy} T. Padmanabhan, Cambridge Univ. Pr., Cambridge, UK, 2010, 700 p.

%\bibitem{hossen} S. Hossenfelder, L. Smolin, Phys. Canada 66 (2010) 99, arXiv:0911.2761 [physics.pop-ph].
%\bibitem{lust} D. Lust, M. Petropoulos, Class. Quant. Grav. 29 (2012) 085013, arXiv:1110.0813 [gr-qc].
\bibitem{magueijo} J. Magueijo, L. Smolin, Class. Quant. Grav. 21 (2004) 1725.
\end{thebibliography}
\end{document}